\documentclass[structabstract]{aa}
\usepackage{natbib}  
\bibpunct{(}{)}{;}{a}{}{,}
\usepackage{graphicx}
\usepackage{txfonts}
\begin{document}
%

    
    \title{Observations of 44 extragalactic radio sources with the VLBA at 92cm:} 
\subtitle{A List of Potential Calibrators and Targets for LOFAR and RadioAstron}

   \author{H. Rampadarath\thanks{\email{rampadarath@jive.nl}} 
          \inst{1,3}
          \and
          M.A. Garrett\thanks{\email{garrett@astron.nl}} 
          \inst{2,3,4}
          \and A. Polatidis\thanks{\email{polatidis@astron.nl} }
          \inst{1,2}
           }
   \institute{Joint Institute for VLBI in Europe (JIVE), Postbus 2, 7990
          AA Dwingeloo, The Netherlands\\              
              \and
          Netherlands Institute for Radio Astronomy (ASTRON), Postbus 2, 7990
           AA Dwingeloo, The Netherlands\\
         \and
             Sterrewacht Leiden, Leiden University, Oort Building, Neils Bhorweg 2, 2333 CA, Leiden,The Netherlands\\      
         \and
           Centre for Supercomputing, Swinburne University of Technology, Mail number H39, P.O. Box 218, Hawthorn, 
           Victoria 3122, Australia\\
             }

   \date{Received February 2, 2009;Accepted March 14, 2009}

 
  \abstract
   {}
   {We have analysed VLBA 92cm archive data of 44 extragalactic sources in order to identify early targets and potential calibrator sources for the LOFAR radio telescope and the RadioAstron space VLBI mission. Some of these sources will also be suitable as ``in-beam'' calibrators, permitting deep, wide-field studies of other faint sources in the same field of view.}
   {All publicly available VLBA 92cm data observed between 1 January 2003 to December 31 2006 have been analysed via an automatic pipeline, implemented within AIPS. The vast majority of the data are unpublished.}
   {The sample consists of 44 sources, 34 of which have been detected on at least one VLBA baseline. 30 sources have sufficient data to be successfully imaged. Most of the sources are compact, with a few showing extended structures. Of the 30 sources imaged, 13 are detected on the longest VLBA baselines ($\sim 9$~M$\lambda$), while all were detected on baselines greater than $2$~M$\lambda$ (the maximum baseline of LOFAR including the current international baselines). }
   {}

\maketitle
%

  \section{Introduction} 

Low-frequency observations of the radio sky are set to be transformed by a new generation of radio telescopes. In particular, instruments such as LOFAR (the Low Frequency Array) will permit large areas of the sky to be surveyed at very low-frequencies (30-230 MHz) with unprecedented sensitivity, and with arcsecond and sub-arcsecond resolution. The base LOFAR configuration foresees at least 36 stations being distributed across the Netherlands with at least 9 additional international stations located in Germany (5), Sweden (1), France (1), the UK (1) and Italy (1). Further stations are planned in other European countries, including Poland, Austria, Ireland and the Ukraine \citep{garetal09}. Initial science observations with LOFAR are expected to begin in 2009. 

The extension of LOFAR to baseline lengths of a few thousand kilometres presents several new challenges. With a resolution of up to $\sim 0.2$ arcseconds, the sky density of suitable calibrator sources may be a significant issue especially at the lowest frequencies. In particular, very little is known about the morphology of radio sources at these low frequencies and high resolutions. A very similar situation exists with the space VLBI mission RadioAstron, due to be launched in October 2009 \citep{zak07}. With a maximum baseline length of 350,000 km, RadioAstron will have a fringe spacing of 0.540 milliarcseconds at 327 MHz (see the RadioAstron booklet for more details\footnote{ http://www.asc.rssi.ru/radioastron}). 

\cite{altetal95} presented the results of a global Mark II VLBI 92 cm snapshot survey. Their sample contained 16 sources, of which there were 3 radio galaxies: 0116+319 (4C31.04), 1117+146 (4C14.41), 2050+364 (DA529); 4 BL Lac objects: 0235+164, 0723-008, 0735+178, 0851+202 (OJ287); and 9 quasars: 0333+321 (NRAO140), 1055+018 (4C01.28), 1422+202, 1611+343 (DA406), 1633+382 (4C38.41), 1901+319 (3C395), 2145+067 (4C06.69), 2230+114 (CTA102), 2251+158 (3C454.3). All of which were resolved on baselines longer than 2-6 M$\lambda$. Only 2 of the sources observed by altetal95 (3C395 and 3C454.3) are part of the sample presented in this paper.

\cite{chupetal99} have presented a list of 61 sources, observed in two separate experiments, at 327 MHz with a global Mark II VLBI array. Of the 61 sources observed, 25 were recommended for further detailed observations and images were made of 2 of these sources. 3 of the sources we present in this paper (3C273, 3C345 and BL Lac), are also included in Chuprikov's sample. 

\cite{lencetal08} have conducted a deep wide field VLBI survey of the sky at 92cm with a global VLBI array using two in-beam calibrators. Of the 272 potentially detectable targets, about 20 sources show compact, milliarcsecond scale structures. These results are encouraging, and suggest that deep, wide-field studies of relatively faint radio sources (using suitable in-beam calibrators) should be a productive way of characterising the low-frequency radio sky at the highest resolution.

All these developments, emphasise the need to establish a suitable list of compact radio sources that may be good calibrators or reasonable commissioning targets for LOFAR and the RadioAstron mission. 
In this paper, we present the results of an automatic analysis of 327 MHz data publicly available via the NRAO Very Long Baseline Array (VLBA), observed during the period 1 January 2003 - 31 December 2006\footnote{The National Radio Astronomy Observatory is a facility of the National Science Foundation operated under cooperative agreement by Associated Universities, Inc}. In Section 2 we discuss the automatic pipeline procedure, and in section 3 we present the main results of our analysis, including images of the individual sources. Notes and brief descriptions of some interesting sources are presented as a separate appendix. Final conclusions are also drawn in section 3.

\section{Data analysis}

All publicly available VLBA 327 MHz data, observed during the period 1 January 2003 - 31 December 2006 were downloaded from the archive and processed by an automatic pipeline implemented in NRAO's AIPS. 23 individual projects were processed (not including many multi-epoch sub-projects), comprising a total of 44 unique radio sources. Table 1 lists the various projects, and the sources observed. 8 of the 23 projects included multiple-pass correlation (see Table 1) to target multiple sources in the same field of view. Table 2 gives details of the individual sources. The sources in the sample are located fairly randomly on the northern sky, with a range of declinations from $\delta$ = -33 degrees to $\delta$ = +77 degrees. The pipeline was designed to amplitude calibrate the data and fringe fit each source individually, using a solution interval of 3 minutes in all cases. The data were averaged in time and frequency, after applying the initial calibration. The pipeline also produced preliminary images of each amplitude calibrated and fringe-fitted (detected) source. For those sources detected, manual editing and imaging was conducted using the AIPS tasks CALIB and IMAGR.

Of the 44 sources, 34 were detected on all or some sub-set of VLBA baselines (table 3) and 10 sources were not detected, on any VLBA baseline: J0440-4333, J0959+6932 (M82), J0955+6940; J1159+0112, J1230-1139, J2007+4029, J2322+0812, J2236+2828, J2334+0736 and J2344+3433. 30 were successfully imaged and 4 sources were either severely resolved or did not have enough data (or UV-coverage) to be successfully imaged. One special case is Cygnus-A. This was barely detected ($9\sigma$) only on the shortest VLBA baseline, and only by introducing an offset point model. For complex sources like Cygnus-A, a more complex model within FRING might result in more detections.

\section{Results and discussion}

Table 3 lists the sources that were detected. Note that most sources were observed by more than 1 project (see table 1). In these cases, the ``best'' observation was selected, based on quality and quantity of the data. Figures 1-5 show the contour plots of the sources \footnote{The corresponding plots of flux density against UV-distance and U vs V plots, will be available on the electronic version}. 
Given that the longest LOFAR baseline at 240 MHz, corresponds to 2M$\lambda$, all sources that are detected on all VLBA baselines are also likely to be detected by LOFAR. All of the 30 sources imaged, are compact enough to be detected on baselines greater than 2 M$\lambda$, and many  show extended structures and are resolved even beyond this baseline length  (e.g. see 3C84, Figure 1). However, for LOFAR, many of these sources will remain unresolved (see the amplitude vs UV-distance plots presented in Figures 1-8). Many of the sources imaged in our programme should therefore make excellent LOFAR primary phase/bandpass calibrators, especially on the longest LOFAR baselines. Table 4 lists the best candidate calibrators. 

Of the 30 imaged sources, only 13 sources have been detected on the very longest VLBA baselines (9-9.5 M$\lambda$). These 13 sources may prove to be interesting targets for RadioAstron. In particular, sources such as 3C84, 1148-001,J2330+11,1510-089,0537-286  and 3C345 may be employed as ``in-beam'' calibrators for deep wide-field 327 MHz ground based VLBI observations. These sources are strong enough to be well detected by ground based VLBI arrays, but not so bright as to significantly raise the system noise of the antennas or introduce potential dynamic range limitations in the image plane.
 
It is interesting to note how the morphology of some of the sources in our sample change with baseline length. A good example is Cygnus-A. The first observations taken with the LOFAR test station at Exloo (1 km baseline) show that Cygnus-A is one of the brightest objects in the sky. However, the VLBA observations at 327 MHz do not detect Cygnus-A on all but the shortest (240 km) baseline (using an offset point source model centred on one of the lobes, see the Appendix for further discussion). We conclude that Cygnus-A will also be heavily resolved for LOFAR on the longest international baselines, at least at the highest observing frequencies. In particular, there is no guarantee that sources that make good calibrators for the LOFAR core, will also be appropriate for longer baseline lengths. Other bright sources will no-doubt show similar properties, although we also note that some of these bright sources (e.g. 3C274) remain bright and compact on milliarcsecond scales. 

More systematic surveys by the VLBA at 327 MHz would be useful to further extend the list of potential calibrators and targets for both LOFAR and RadioAstron. 

\section*{APPENDIX: Individual source comments}

In this Appendix the properties of some of the more interesting sources in our sample are discussed.

\subsection*{J0405+3803}
{The data presented in this paper for J0405+3803, have previously been observed and published by \cite{rodetal06}. The images of 0402+379 produced in this article, compare well with those of \cite{rodetal06}, and are consistent with lower resolution VLA images taken at 1.4 GHz \citep{xuetal95}.

\subsection*{J0433+0521 (3C120)}
The VLBA images at 327 MHz presented here are consistent with the images at 2.3 GHz taken by \cite{fey&char00}, also using the VLBA. The results at 2.3 GHz presents the core and jet of 3C120 with 3 jet components, all located within approximately 100 mas of the core. The results presented here show an extended jet that spans $\sin$ 300 mas.

\subsection*{J0542+4951 (3C147)}
{3C147 is very resolved at 327 MHz and shows an elongated structure that spans $\sim 400$ milliarcseconds, largely consistent with that observed by EVN-MERLIN at 1.6 GHz \citep{davisetal96}. The image is also in good agreement with higher resolution VLBA observations made at 8.5 GHz and 2.3 GHz \citep{fey&char00}. We do not detect the emission seen by MERLIN to the North-East of the peak emission, that \citep{davisetal96} interpret as evidence of outflow from a counter jet. There is als0 an extended structure to the South-East of the peak of the emission, which is also observed at 1.6 GHz with EVN-MERLIN and, 2.3/8.5 GHz (VLBA).} 

\subsection*{J1230+1223 (M87, 3C274)}
The observations with the VLBA at 327 MHz show that the well-known jet extends up to $\sim 500$ mas (0.5 arcsecs) from the core. The superluminal component observed by the Hubble Space Telescope known as HST 1, \citep{bireetal99} is clearly detected, located approximately 0.8 arcsecs from the core. This has also recently been detected by the VLBA at 15 GHz \citep{changetal08}. At 15 GHz HST-1 appears to be variable. When detected it has a flux density in the range of 14-22 mJy.  At 327 MHz, we measure an integrated flux density of 317 mJy, implying a spectral index of $\alpha_{0.327}^{15} \sim -0.697$. The size of the component at 327 MHz is $48 \times 21$ milliarcseconds in PA $\sim 144$ degrees.

\subsection*{J1331+3030 (3C286)} 
{Our images of 3C286 are consistent with other VLBA images at higher frequencies e.g. \cite{fey&char00}. The 327 MHz observations are clearly extended in the same direction as that seen in the MERLIN images at 18 cm \citep{akuetal94}. We do not see the counter jet report by \citep{akuetal94}.}

\subsection*{J1512-0905}
{The image presented here at 327 MHz shows a jet component to the North-East of the core, which ends in a hot-spot.  At higher frequencies (2.3, 8.6 and 15 GHz), the hot-spot is not visible by the VLBA, but there is evidence for a jet component along the direction of the hot-spot \citep{fey&char97,lister&homan05}.

\subsection*{J1746+6226 (4C +62.29)}
{At 327 MHz, we observe an extended core component with a morphology that is consistent with higher frequency VLBA observations at 2.3 GHz and 8.6 GHz \cite{fey&char00}. However, our images also reveal a counter jet with a hot-spot located approximately 0.2 arcseconds to the North-East of the extended core. The PA of this counter jet feature is diametrically opposite to the higher frequency jet that is seen on the other side of the core.}

\subsection*{J1829+4844 (3C380)}
{The radio source J1829+4844 is identified with the very powerful steep spectrum radio source, 3C380. Observations at decimetre and centimetre wavelengths have identified this object as an FR II source that is elongated along the line-of-sight \citep{wilketal91,meganetal06}. This source has a very extended structure, with a number of compact components that corresponds to the core coincident with the optical quasar with a short (15 mas) jet, a hot-spot in the nearer lobe 0.7 arcseconds from the core, and another slightly larger hot-spot 0.4 arcseconds from the first \citep{meganetal06}. The lobes are located to the North-West of the core. Only the core and the hot-spot at 0.7 arcseconds from the core are visible in the VLBA observations at 327 MHz. }

\subsection*{J1902+3159 (3C395)}
{3C395 is identified with a 17th magnitude quasar \citep{geld&whit94}, and has a complex radio structure on a wide range of angular scales. At 8.3 GHz with the VLBA its core-jet radio structure consists of two components (core and a hot-spot) separated by approximately 15 mas, with the hot-spot located South-West of the core \citep{laraetal97}. VLBI images at 326 MHz, taken by \cite{altetal95} show a compact and unresolved core, and an extended regio of emission separated by approximately 65 mas South-West of the core. Our results agree well with these low resolution images. There is also a fainter, extended component observed in our image at P.A. 111.7$^{\circ}$}. This was not observed by \cite{altetal95}. 

\subsection*{J1959+4044 (Cygnus-A)}
Despite being one of the brightest sources in the sky, Cygnus A was not detected by the VLBA at 327 MHz, except on the shortest baselines after including an offset point source model for the FRING task - offset in the direction of the South-Eastern lobe. Since it is located at low galactic latitude (b = 5 degrees), the core of Cygnus-A may be scatter broadened due to interstellar scattering in our own galaxy \citep{cari91}. With a measured size of 1.6 mas at 5 GHz, we would expect the core size to be $\sim 400 $ mas at 327 MHz. The core of Cygnus-A is therefore likely to be resolved by our observations. The same may also be true of hot-spots in the two lobe structures. 

\begin{acknowledgements}
This research was supported by the EU Framework 6 Marie Curie Early Stage Training programme under contract number MEST-CT-2005-19669 "ESTRELA".
\end{acknowledgements}

\bibliographystyle{aa}
\bibliography{1802}


\onecolumn
\begin{table}
\caption{327 MHz VLBA archived projects from 2003 to 2006}
\label{table:1}
\begin{tabular}{l l l}
\hline                   

Project & Date of & Common Source name \\
		& Observation & \\

\hline                                
BM201B & 16/07/03 & 3C454.3, J2327+0940, J2334+0736, J2336+2828 \\
BM193 PASS1 & 30/08/03 & NGC 7674, J2330+1100, 3C454.3 \\
BM193 PASS2 & 30/08/03 & J2322+081, J2330+11, 3C454.3 \\
BC138A PASS1 & 09/08/03 & J1331+3030, 3C336, J1816+3457 \\
BC138A PASS2 & 09/08/03 & J1331+3030, J1619+2247, J1816+3457 \\
BEO32A PASS1 & 09/11/03 & 3C286, 3C395, 3C418, 3C454.3, BLLAC, Cygnus-A, J1821+3945, J1829+4844, J1924+3329, J2007+4029 \\
BEO32A PASS2 & 09/11/03 & 3C286, 3C395, 3C418, 3C454.3, BLLAC, J2007+4029, J1821+3945, J1829+4844, J1924+3329, J2007+4029 \\
BWO67Q & 31/12/03 & 3C120, 3C286, 3C147, J0423-0120 \\
BC138C & 30/01/04 & 3C273, 3C275.1, J1331+3030 \\
BH116 & 11/03/04 & 3C286;J1512-0905 \\
BJ046R & 23/04/04 & J1150-0024, 3C147, 3C274 \\
BB197 & 08/07/04 & 3C454.3, J2022+6136, J1829+4844 \\
BH126A & 23/12/04 & 3C147, J1150-002 \\
BC150  & 11/02/05 & J1746+6226, 3C286, 3C345 \\
BT070B & 13/06/05 & J0405+3803, 3C111, 3C84 \\
BFO87 PASS1 & 03/12/05 & J2005+7752, J0955+6940 \\
BF087 PASS2 & 03/12/05 & M82, J2005+7752 \\
BH135D & 06/02/06 & J0955+6940, J1150-002, 3C147, 3C274 \\ 
BK131C  & 08/06/06 & J1230-1139, J1333+1649, 3C286, 3C345, J0539-283 \\
BK131A & 09/06/06 & J0440-4333, J0204+3649,J0319+4130, J2344+3433, J0405+3803, 3C454.3 \\
BK131B & 18/06/06 & J0440-4333, J1159+0112, 3C286, J0539-2839, J2344+3433, 3C454.3 \\
BH135G & 03/07/06 & J1150-0024, 3C147, 3C274 \\

\hline

\end{tabular}


\caption{General properties of the 44 radio sources detected by VLBA at 92cm from 2003-2006. \textit{Obtained from The NASA Extragalactic Database}}
\label{table:2}

\begin{tabular*}{\textwidth}{@{\extracolsep{\fill}}p{0.17\textwidth}p{0.15\textwidth}p{0.1\textwidth}p{0.1\textwidth}p{0.15\textwidth}p{0.17\textwidth}}

\hline

IAU J2000 Source Name & Other Aliases & Type & Redshift & R.A. (J2000) & Decl. (J2000) \\
\hline

J0204+3649 & 0201+365 & Q & 2.912000 & 02h04m55.596s &  +36d49m17.996s \\ [-0.1ex]
J0319+4130 & 3C84 & G & 0.017559 & 03h19m48.160s & +41d30m42.103s \\
J0405+3803 & 4C+37.11& G & 0.055000 & 04h05m49.262s & +38d03m32.236s \\
J0407-3303 & 0405-331 & Q & 2.562000 & 04h07m33.914s &-33d03m46.359s \\
J0418+3801 & 3C111 & G & 0.048500 & 04h18m21.277s & +38d01m35.900s \\
J0423-0120 & - & Q & 0.914000 & 04h23m15.801s &-01d20m33.065s \\
J0433+0521 & 3C120 & G & 0.033100 & 04h33m11.096s & +05d21m15.619s \\
J0440-4333 & 0438-436 &  Q & 2.863000 & 04h40m17.180s & -43d33m08.604s \\
J0542+4951 & 3C147 & Q & 0.545000 & 05h42m36.138s & +49d51m07.234s \\
J0539-2839 & 0537-286 & Q & 3.104000 & 05h39m54.281s & -28d39m55.948s \\
J0959+6932 & M82 & G & 0.000677 & 09h59m10.639s  &  +69d32m17.724s \\
J0955+6940 & - & - & - & 09h59m55.694 & +69d40m43.690s \\
J1150-0024 & 1148-001 & Q & 1.976200 & 11h50m43.871s & -00d23m54.205s  \\
J1230+1223 & 3C274, M87 &  G  & 0.004360 & 12h30m49.423s &  +12d23m28.044s \\
J1159+0112 & 1157+014 & Q & 1.999650 & 11h59m44.829s & +01d12m06.960s \\
J1229+0203 &  3C273 & Q &  0.158339  &  12h29m06.700s  &  +02d03m08.598s \\
J1230-1139 & 1228-113 & Q & 3.528000 & 12h30m55.556s & -11d39m09.796s \\
J1243+1622 & 3C275.1  & Q & 0.555100  & 12h43m57.657s &  +16d22m53.440s \\
J1331+3030 &  3C286 & G  & 0.586000 & 13h31m08.288s & +30d30m32.959s \\
J1333+1649 & 1331+170 & Q & 2.084000 & 13h33m35.783s & +16d49m04.015s \\
J1512-0905 & 1510-089 & Q &  0.360000  &  15h12m50.533s & -09d05m59.829s \\
J1619+2247 & - & Q & 1.987000 & 16h19m14.825s & +22d47m47.85s \\
J1624+2345 & 3C336 & Q & 0.927398 & 16h24m39.088s  &  +23d45m12.240s \\
J1642+3948 & 3C345  & Q  &  0.592800  &  16h42m58.810s &  +39d48m36.994s \\
J1746+6226 & 4C+62.29 &  Q &  3.889000  & 17h46m14.034s  & +62d26m54.738s \\
J1816+3457 & - & G  &  0.244800 & 18h16m23.901s  &  +34d57m45.748s \\
J1821+3945 & - &  RS & - & 18h21m59.701s  & +39d45m59.657s \\
J1829+4844 & 3C380 & Q & 0.692000 & 18h29m31.739s & +48d44m46.971s \\
J1902+3159 & 3C395 & Q  &  0.635000  & 19h02m55.939s & +31d59m41.702s \\
J1924+3329 & 4C+33.48  &  RS  & - &  19h24m17.476s  &  +33d29m29.484s \\
J1959+4044 & Cygnus-A &  Q &  0.056075  & 19h59m28.357s & +40d44m02.097s \\
J2005+7752 & 2007+777 & Q &  0.342000 &  20h05m30.999s  &  +77d52m43.248s \\
J2007+4029 & 2005+403  &  Q & 1.736000 &  20h07m44.945s &  +40d29m48.604s \\
J2022+6136 & - & G  & 0.227000 & 20h22m06.682s & +61d36m58.805s \\
J2038+5119 & 3C418 & Q & 1.686000  & 20h38m37.035s  & +51d19m12.663s \\
J2202+4216 & BL Lac  & Q &  0.068600  & 22h02m43.291s &  +42d16m39.980s \\
J2253+1608 & 3C454.3 & Q & 0.859000 & 22h53m57.748s & +16d08m53.561s \\
J2322+0812 & 2320+079 & Q & 2.090000 & 23h22m36.089s & +08d12m01.593s \\
J2327+0846 & NGC7674 & G & 0.028924 & 23h27m56.710s & +08d46m44.135s \\
J2327+0940 & - & Q & 1.843000 & 23h27m33.581s & +09d40m09.463s \\
J2330+1100 & - & Q & 1.489000 & 23h30m40.852s & +11d00m18.710s \\
J2334+0736 & - & Q & 0401000 & 23h34m12.828s & +07d36m27.552s \\
J2336+2828 & - & RS & - & 23h36m22.471s & +28d28m57.413s \\
J2344+3433 & 2342+342 & Q & 3.053000 & 23h44m51.254s & +34d33m48.640s \\
\hline

\hline

\end{tabular*}
\end{table}

\begin{table}

\caption{List of 34 detected sources. The images of most of these sources are shown in Figures 1-5.}
\bigskip
\label{table:3}
\begin{tabular}{lcccc}
\hline
 Source & Detected  Baseline (M$\lambda$) & Imaged (Y/N) & Peak Flux Density (Jy/beam) & Total Flux Density, $S_{327}$ (Jy) \\
 \hline

J0204+3649 & 8.0 & Y & 0.337 & 0.481\\
J0319+4130 (3C84) & 9.0 & Y & 2.289 & 4.625  \\
J0405+3803 & 4.0 & Y & 0.427 & 0.539 \\
J0407-3303 & 9.5 & Y & 0.200 & 0.225 \\
J0418+3801 (3C111) & 6.0 & Y & 0.328 & 0.533 \\
J0423-0120 & 9.5 & Y & 0.869 & 1.099 \\
J0433+0521 (3C120) & 9.5 & Y & 1.263 & 1.771 \\
J0542+4951 (3C147) & 8.0 & Y & 3.850 & 40.980 \\
J0539-2839 & 9.5 & Y & 0.498 & 0.564 \\
J1150-0024 & 9.5 & Y & 2.645 & 3.403 \\
J1230+1223 (3C274) & 9.5 & Y & 1.199 & 2.154 \\
J1229+0203 (3C273) & 9.5 & Y & 4.028 & 5.23 \\
J1243+1622 (3C275.1) & 2.0 & Y & 0.422 & 1.413\\
J1331+3030 (3C286) & 9.5 & Y & 6.289 & 16.926   \\
J1333+1649 & 6.0 & Y & 0.043 & 0.474 \\
J1512-0905 & 9.5 & Y & 0.480 & 0.611  \\
J1619+2247 & 2.5 & N & - & - \\
J1624+2345 (3C336) &  4.5 &  N & - & - \\
J1642+3948 (3C345) & 9.5 & Y & 2.952 & 3.492 \\
J1746+6226 & 6.5 & Y & 0.038 & 0.131 \\
J1816+3457 & 6.0 & Y & 0.440 & 0.729  \\
J1821+3945 & 5.0 & Y & 1.748 & 3.974 \\
J1829+4844 & 8.0 & Y & 2.948 & 3.880 \\
J1902+3159 (3C395) & 5.0 & Y & 0.842 & 2.100 \\
J1924+3329 & 2.0 & Y & 0.826 & 0.940  \\
J1959+4044 (Cygnus-A) & 0.3 & N & - & - \\
J2005+7752 & 6.0 & Y & 0.508 & 0.602  \\
J2022+6136 & 9.5 & Y & 0.635 & 0.990 \\
J2038+5119 (3C418) & 4.0 & Y & 1.234 & 3.320 \\
J2202+4216 (BL Lac) & 6.0 & Y & 0.938 & 1.798 \\
J2253+1608 (3C454.3) & 9.0 & Y & 5.300 & 8.659  \\
J2327+0846 (N7674) & 1.8 & N & - & - \\
J2327+0940 & 7.0 & Y & 0.317 & 0.396 \\
J2330+1100 & 9.0 & Y & 0.774 & 0.844 \\

\hline
\end{tabular}
\end{table}


\begin{table}
\caption{Best primary candidate calibrators for LOFAR on baselines $\backsim$ 2M$\lambda$.}
\label{table:4}
\begin{tabular}{lccccc}
\hline
Source & Type & Redshift & R.A. & Dec. & $S_{327}$ (Jy) \\
\hline

J0319+4130 (3C84) & Q & 0.017559 & 03h19m48.160s & +41d30m42.103s & 4.625 \\
J0405+3803 & G & 0.055000 & 04h05m49.262s & +38d03m32.236s & 0.539\\
J0418+3801 (3C111) &  G & 0.048500 & 04h18m21.277s & +38d01m35.900s & 0.533\\
J0423-0120 &  Q & 0.914000 & 04h23m15.801s & -01d20m33.065s & 1.099\\
J0539-2839 &  Q & 3.104000 & 05h39m54.281s & -28d39m55.948s & 0.564\\
J1150-0024 & Q & 1.976200 & 11h50m43.871s & -00d23m54.205s  & 3.403\\
J1642+3948 (3C345)  &  Q  &  0.592800  &  16h42m58.810s &  +39d48m36.994s & 3.492\\
J1816+3457 &  G  &  0.244800 & 18h16m23.901s  &  +34d57m45.748s & 0.729\\
J2005+7752 &  Q  &  0.342000 &  20h05m30.999s  &  +77d52m43.248s & 0.602 \\
J2022+6136 &  G  & 0.227000 & 20h22m06.682s & +61d36m58.805s & 0.990\\
J2202+4216 (BL Lac)  &   Q &  0.068600  & 22h02m43.291s &  +42d16m39.980s & 1.798\\
J2253+1608 (3C454.3) & Q & 0.859000 & 22h53m57.748s & +16d08m53.561s & 8.659 \\
J2327+0940 & Q & 1.843000 & 23h27m33.581s & +09d40m09.463s & 0.396\\
J2330+1100 & Q &  1.489000 & 23h30m40.852s & +11d00m18.710s & 0.844\\

\hline
\end{tabular}
\end{table}

\pagebreak 
\begin{figure}

\centering
\caption{Imaged sources at 327 MHz. The sources here in order are: J0204+3649, J0319+4130 (3C84), J0405+3803, J0407-3303, J0418+3801 (3C111), J0423-0120}
\includegraphics[scale=0.9]{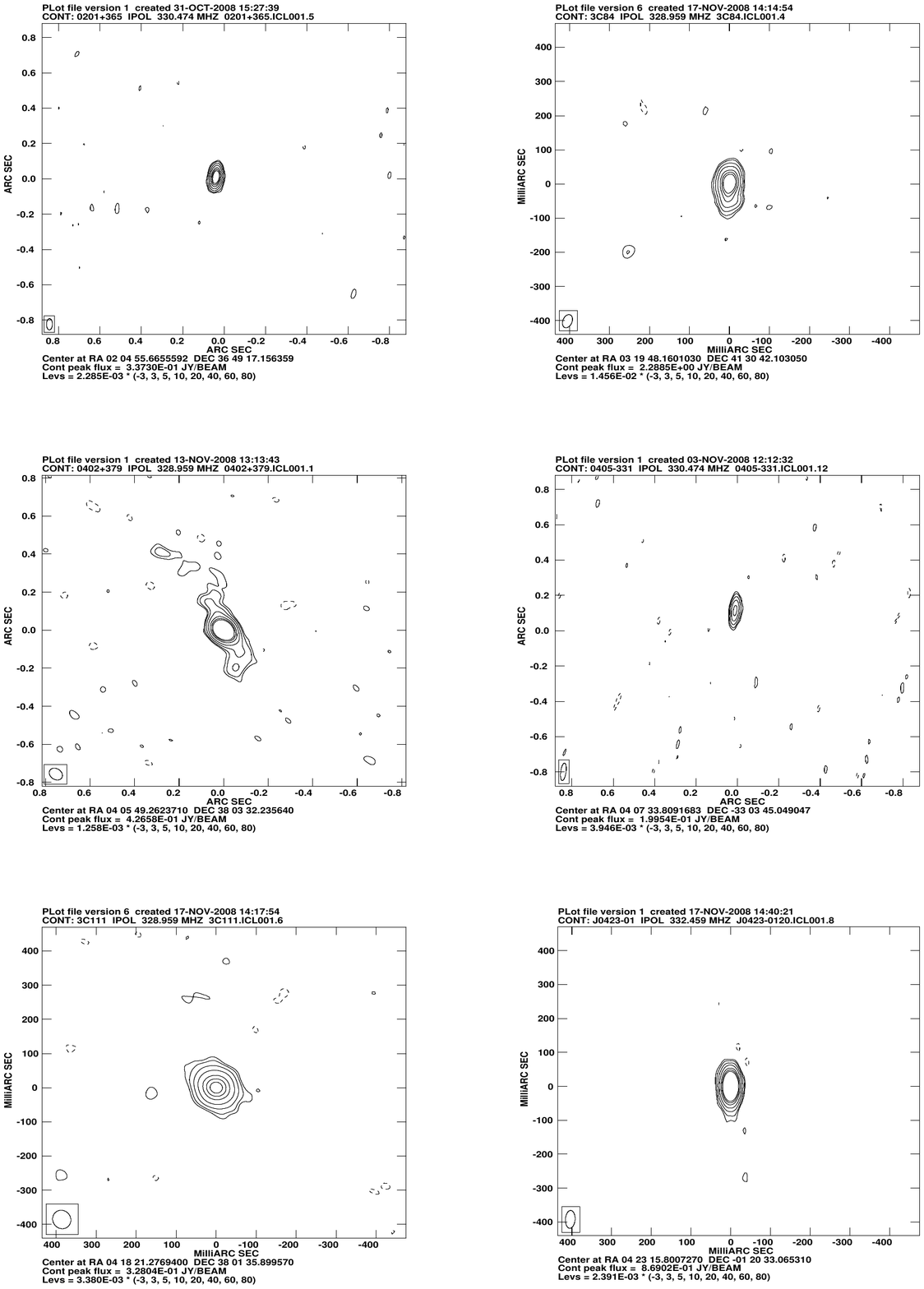}   
\end{figure}

\begin{figure}
\pagebreak 
\caption{Imaged sources at 327 MHz. The sources here in order are: J0433+0521 (3C120), J0542+4951 (3C147), J0539-2839, J1150-0024, J1230+1223 (3C274), J1229+0203 (3C273)}
\includegraphics[scale=0.9]{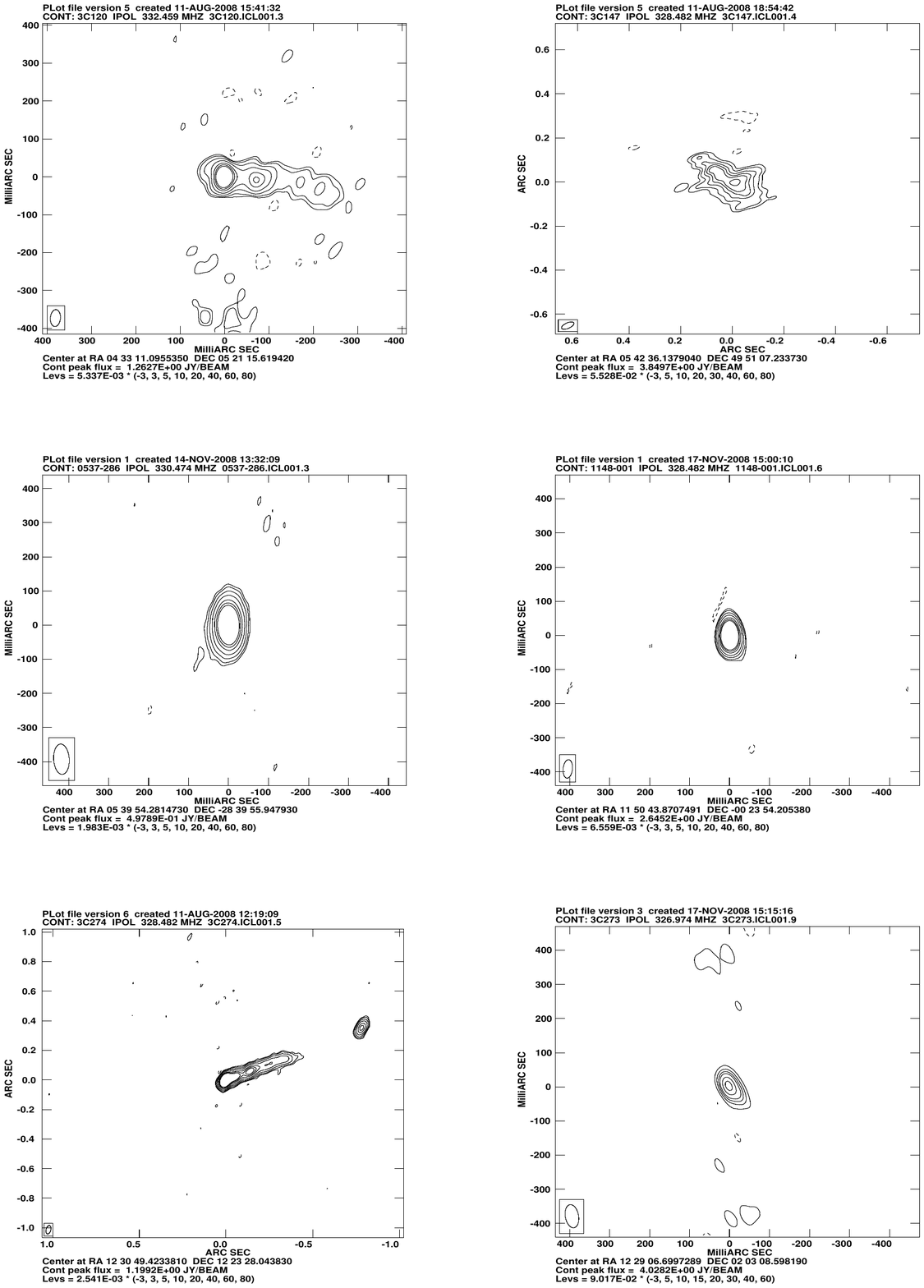}   
\end{figure}

\begin{figure}
\pagebreak 
\caption{Imaged sources at 327 MHz. The sources here in order are: J1243+1622 (3C275.1), J1331+3030 (3C286), J1333+1649, J1512-0905, J1642+3948 (3C345), J1746+6226}
\includegraphics[scale=0.9]{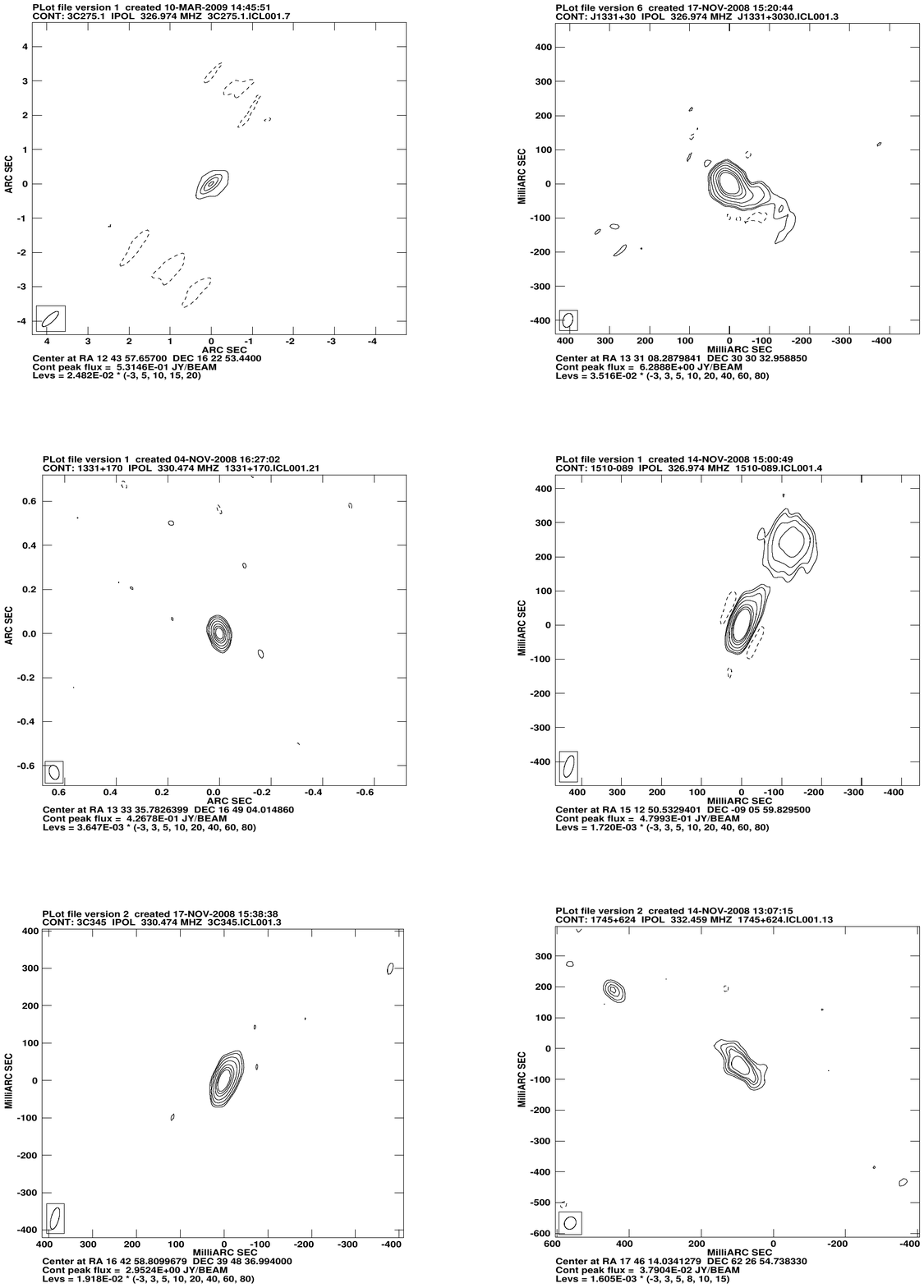}   
\end{figure}

\begin{figure}
\pagebreak 
\caption{Imaged sources at 327 MHz. The sources here in order are: J1816+3457, J1821+3945, J1829+4844, J1902+3159 (3C395), J1924+3329, J2005+7752}
\includegraphics[scale=0.9]{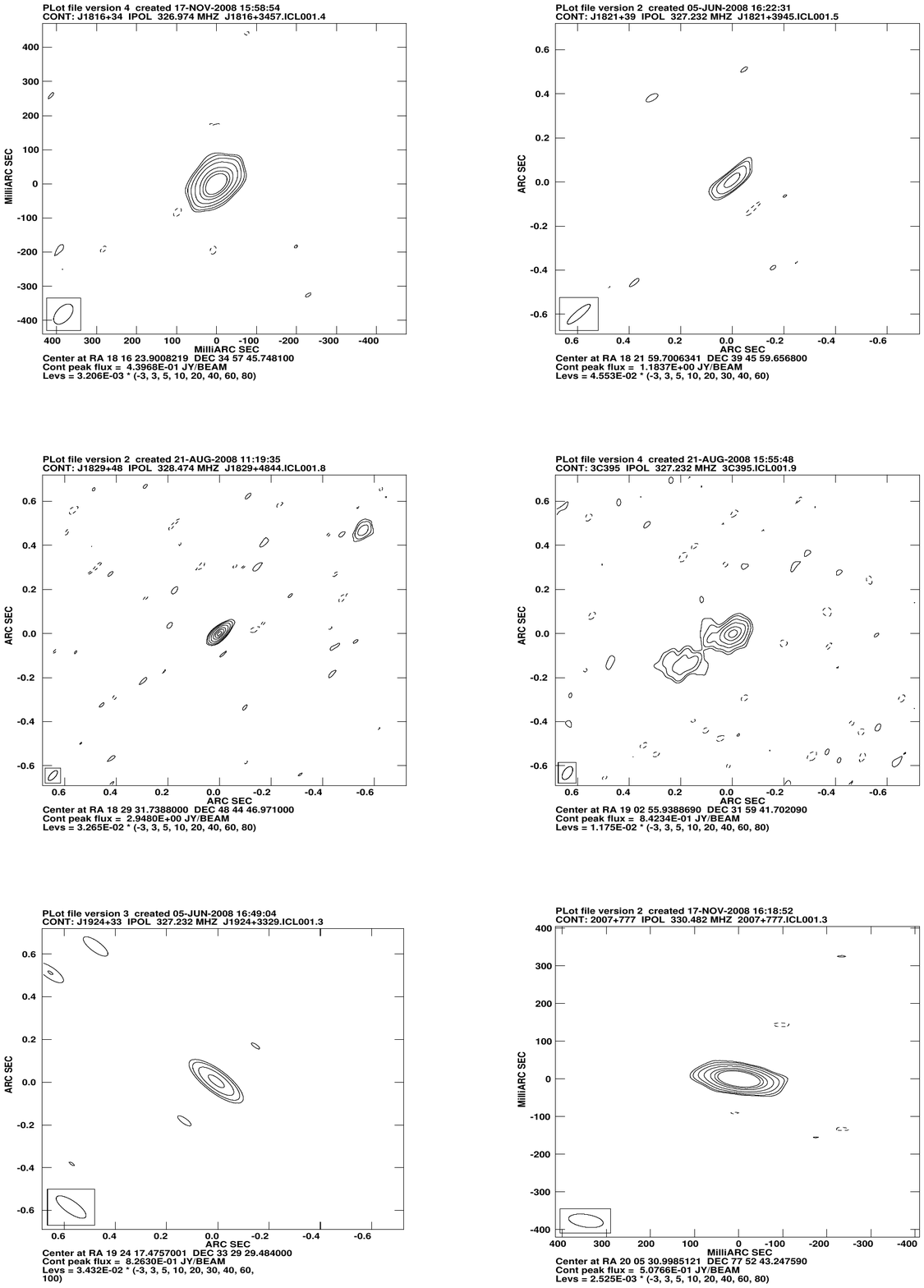}   
\end{figure}

\begin{figure}
\pagebreak 
\caption{Imaged sources at 327 MHz. The sources here in order are: J2022+6136, J2038+5119 (3C418), J2202+4216 (BL Lac), J2253+1608 (3C454.3), J2327+0940, J2330+11}
\includegraphics[scale=0.9]{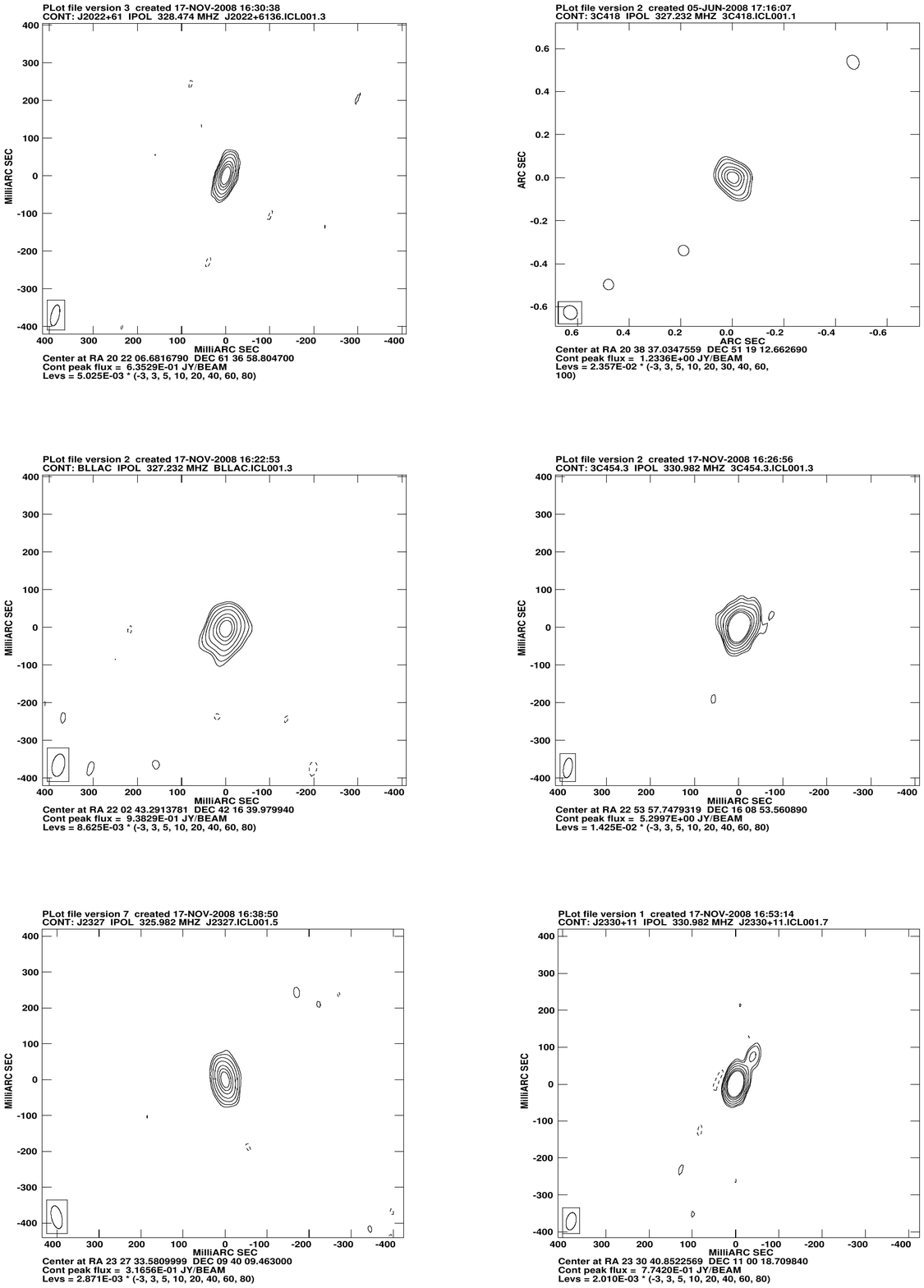}   
\end{figure}

\begin{figure}
\caption{Plots of flux density vs UV distance and UV plots for J0204+3649, J0319+4130 (3C84), J0405+3803}
\includegraphics[scale=0.9]{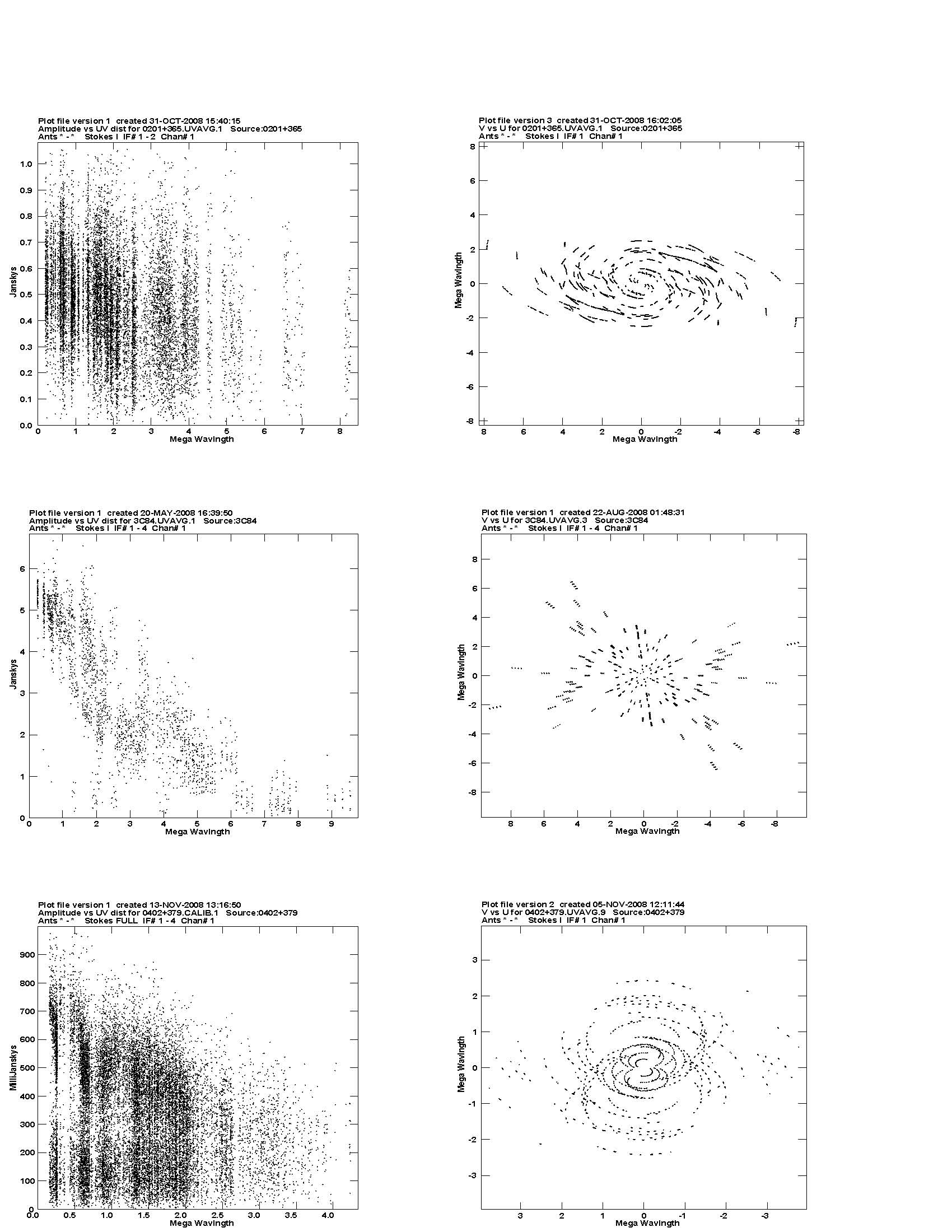} 
\end{figure}

\begin{figure}
\caption{Plots of flux density vs UV distance and UV plots for J0407-3303, J0418+3801 (3C111), J0423-0120} 
\includegraphics[scale=0.9]{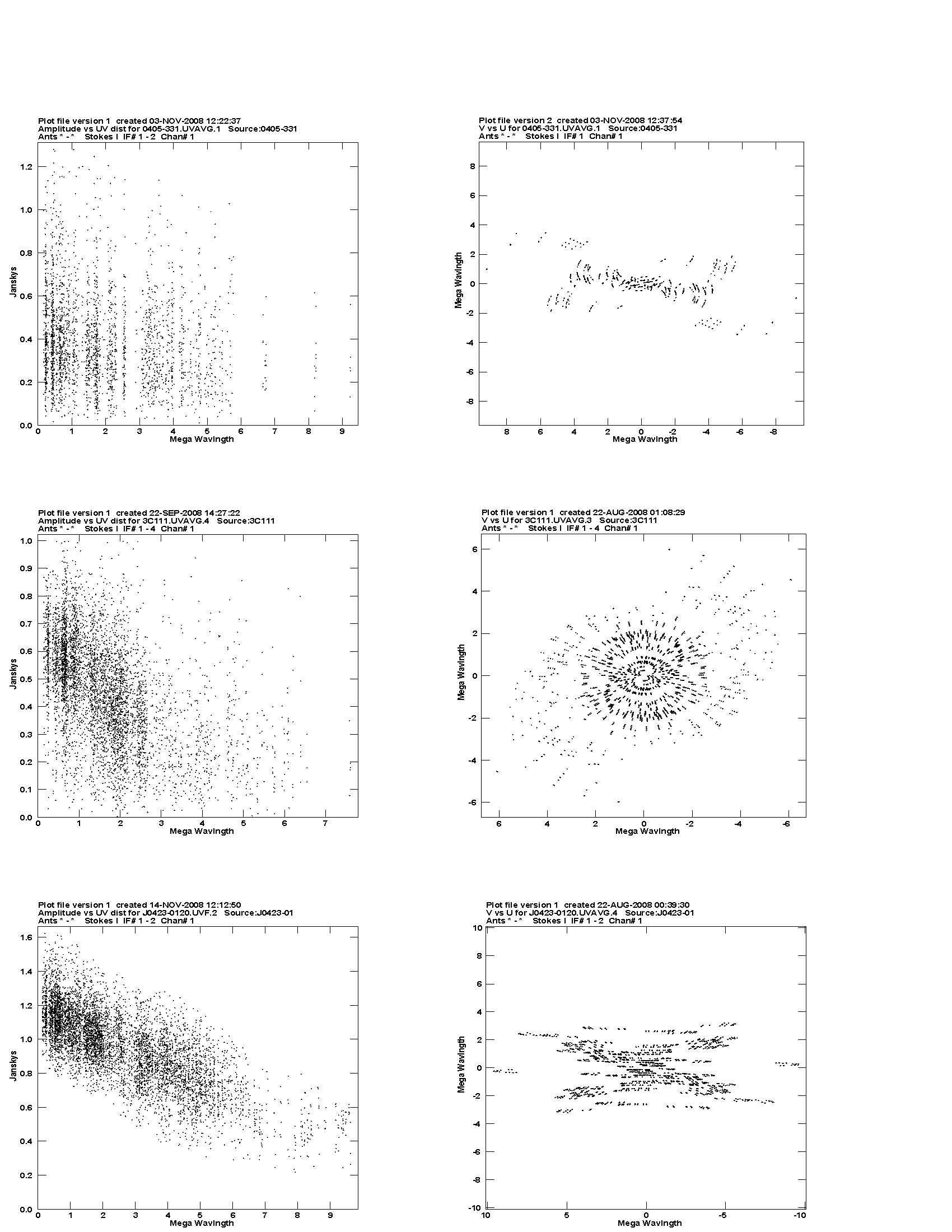} 
\end{figure}

\begin{figure}
\caption{Plots of flux density vs UV distance and UV plots for J0433+0521 (3C120), J0542+4951 (3C147), J0539-2839} 
\includegraphics[scale=0.9]{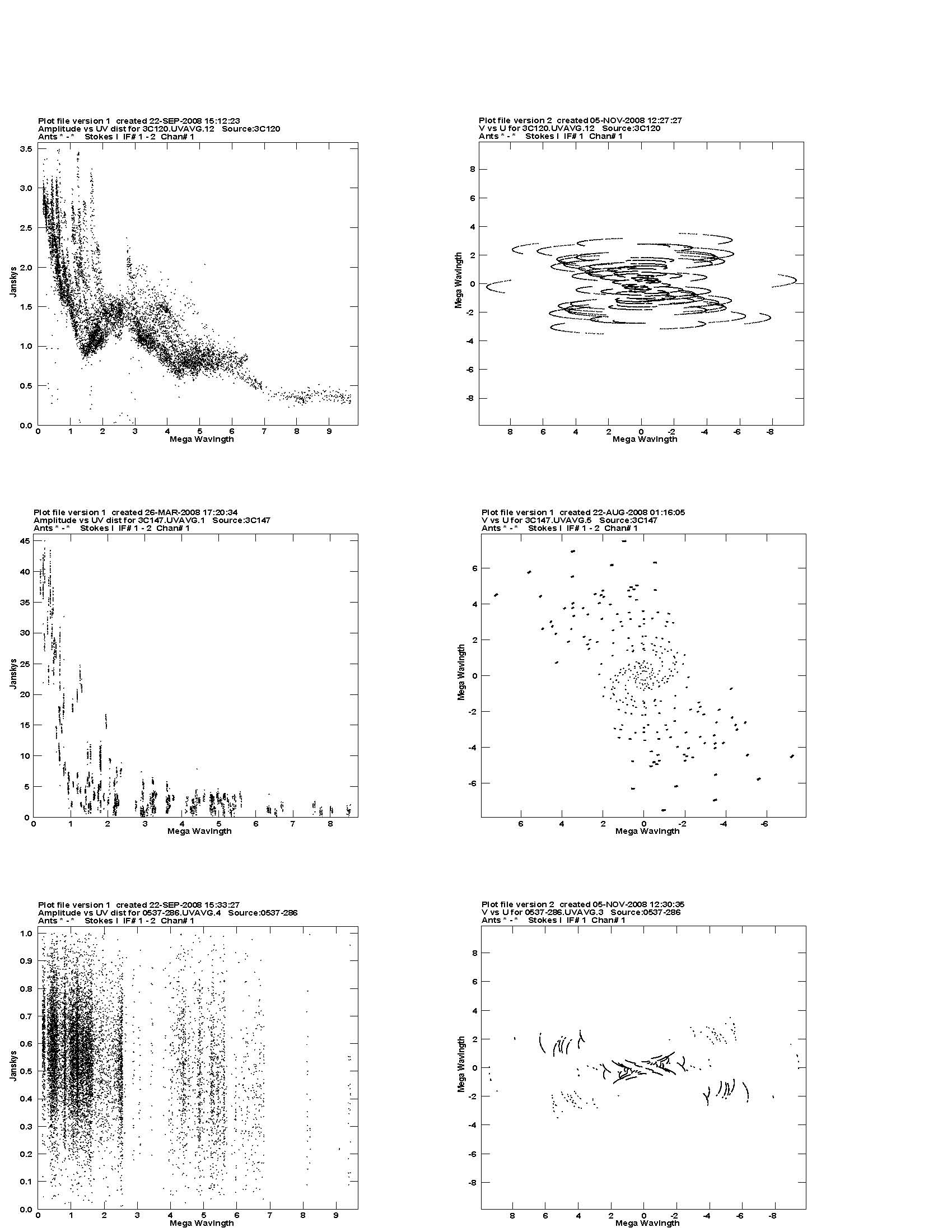} 
\end{figure}

\begin{figure}
\caption{Plots of flux density vs UV distance and UV plots for J1150-0024, J1230+1223 (3C274), J1229+0203 (3C273)} 
\includegraphics[scale=0.9]{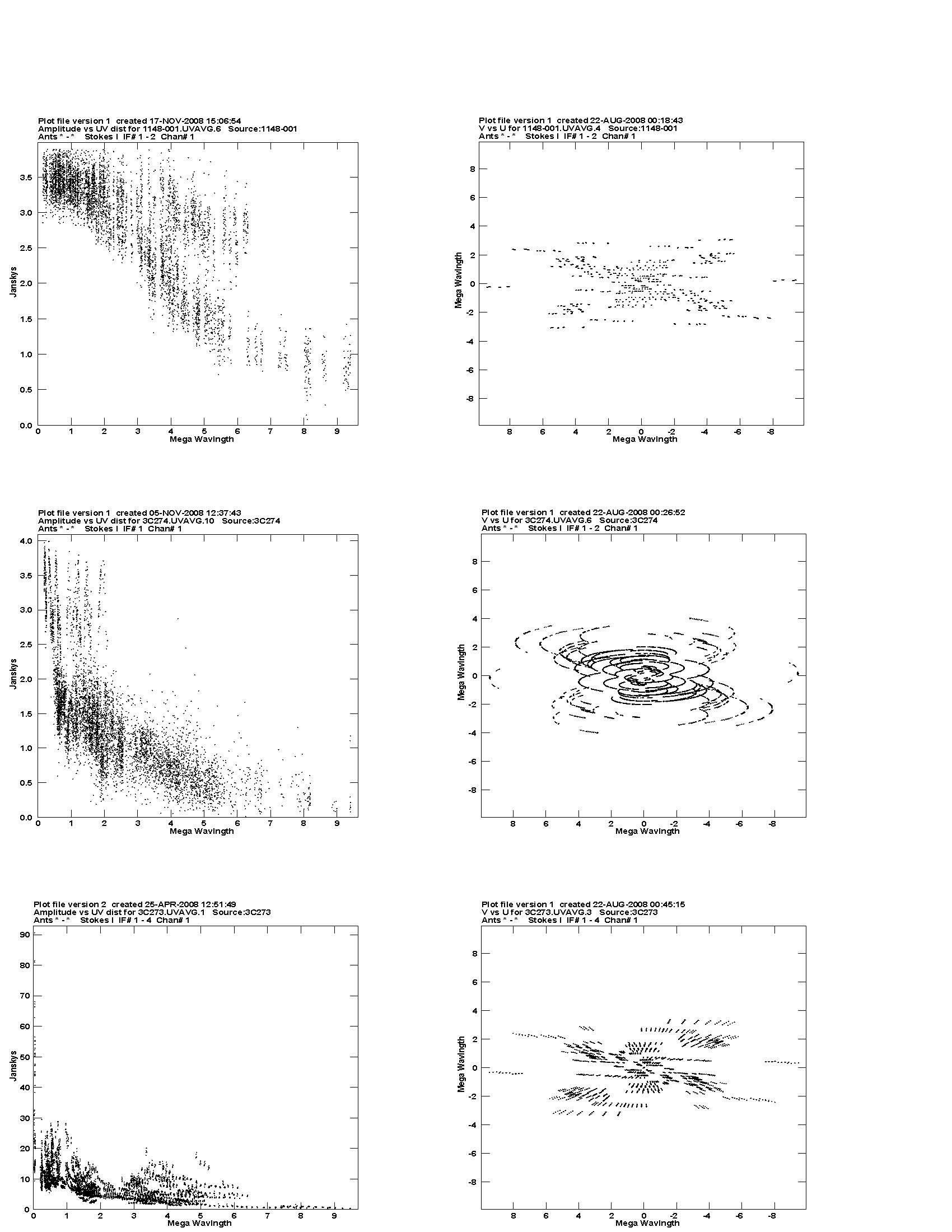} 
\end{figure}

\begin{figure}
\caption{Plots of flux density vs UV distance and UV plots for J1243+1622 (3C275.1), J1331+3030 (3C286), J1333+1649} 
\includegraphics[scale=0.9]{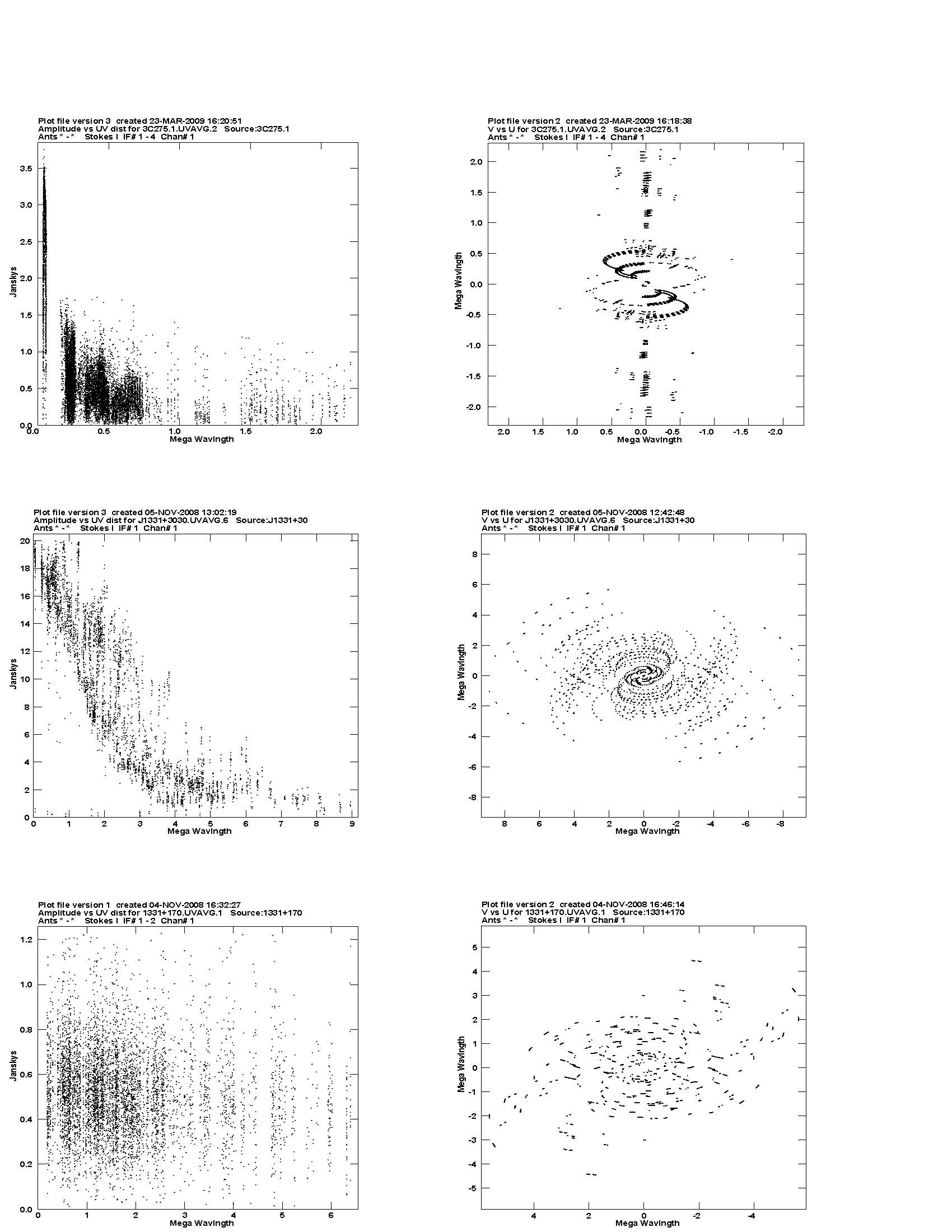} 
\end{figure}

\begin{figure}
\caption{Plots of flux density vs UV distance and UV plots for J1512-0905, J1642+3948 (3C345), J1746+6226} 
\includegraphics[scale=0.9]{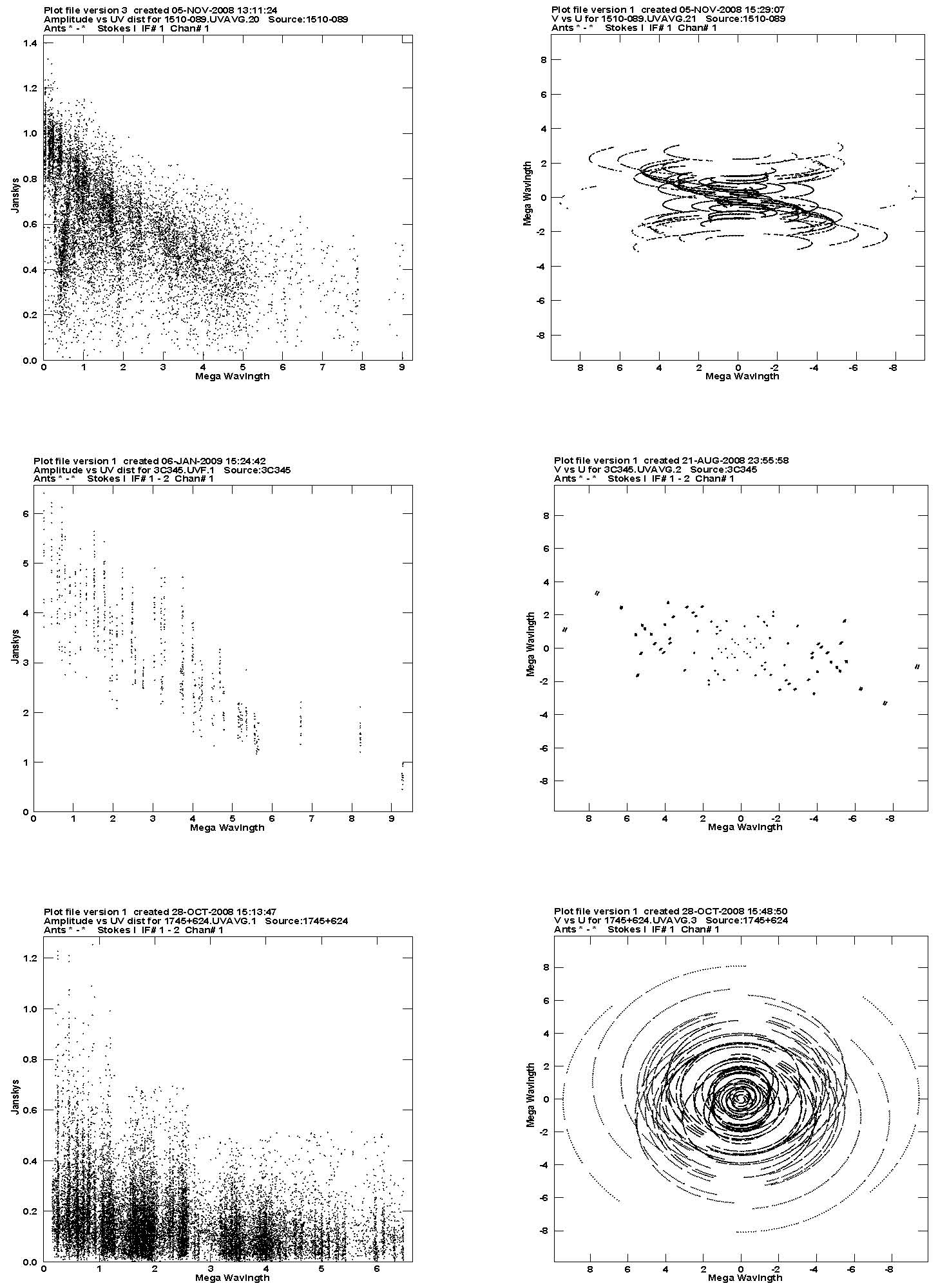} 
\end{figure}

\begin{figure}
\caption{Plots of flux density vs UV distance and UV plots for J1816+3457, J1821+3945, J1829+4844} 
\includegraphics[scale=0.9]{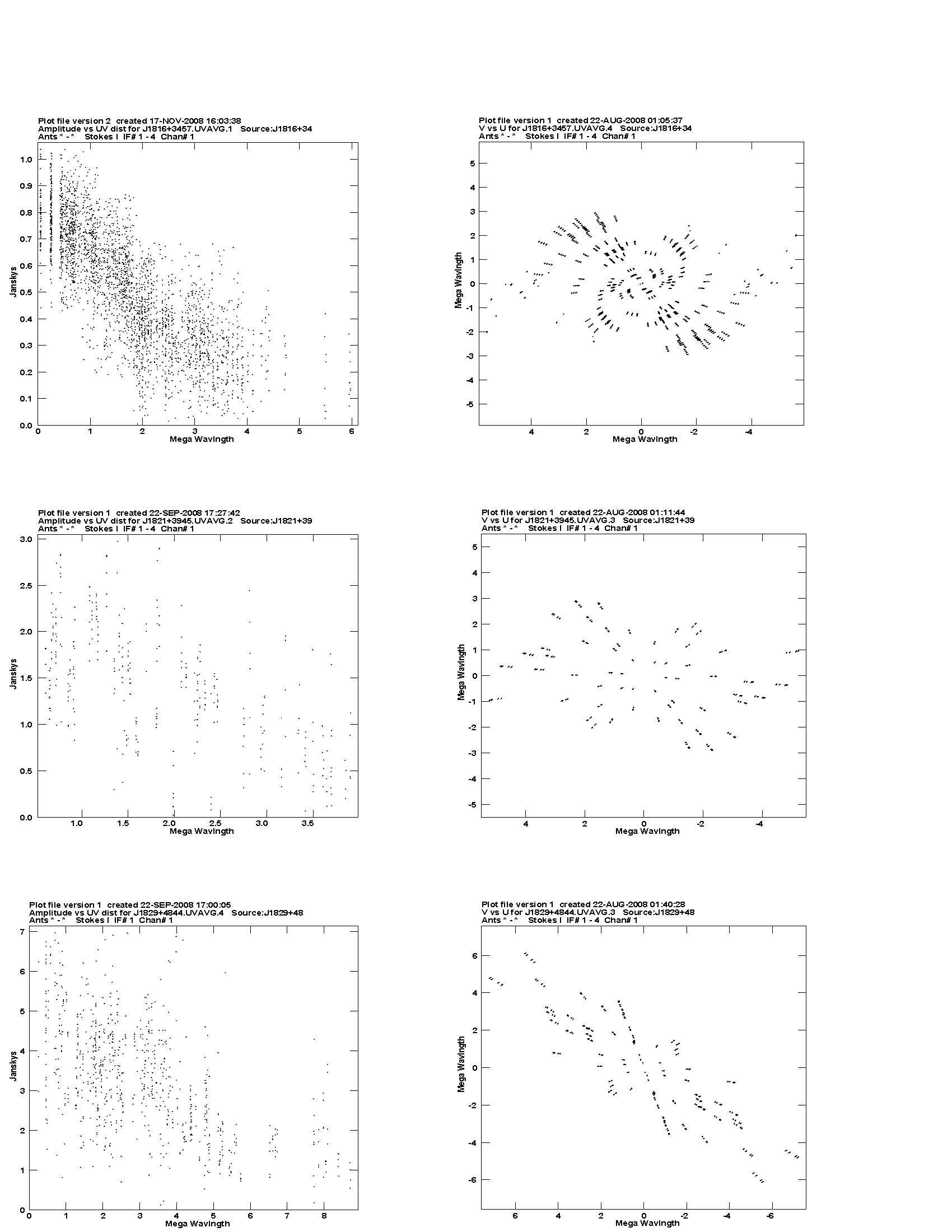} 
\end{figure}

\begin{figure}
\caption{Plots of flux density vs UV distance and UV plots for J1902+3159 (3C395), J1924+3329, J2005+7752} 
\includegraphics[scale=0.9]{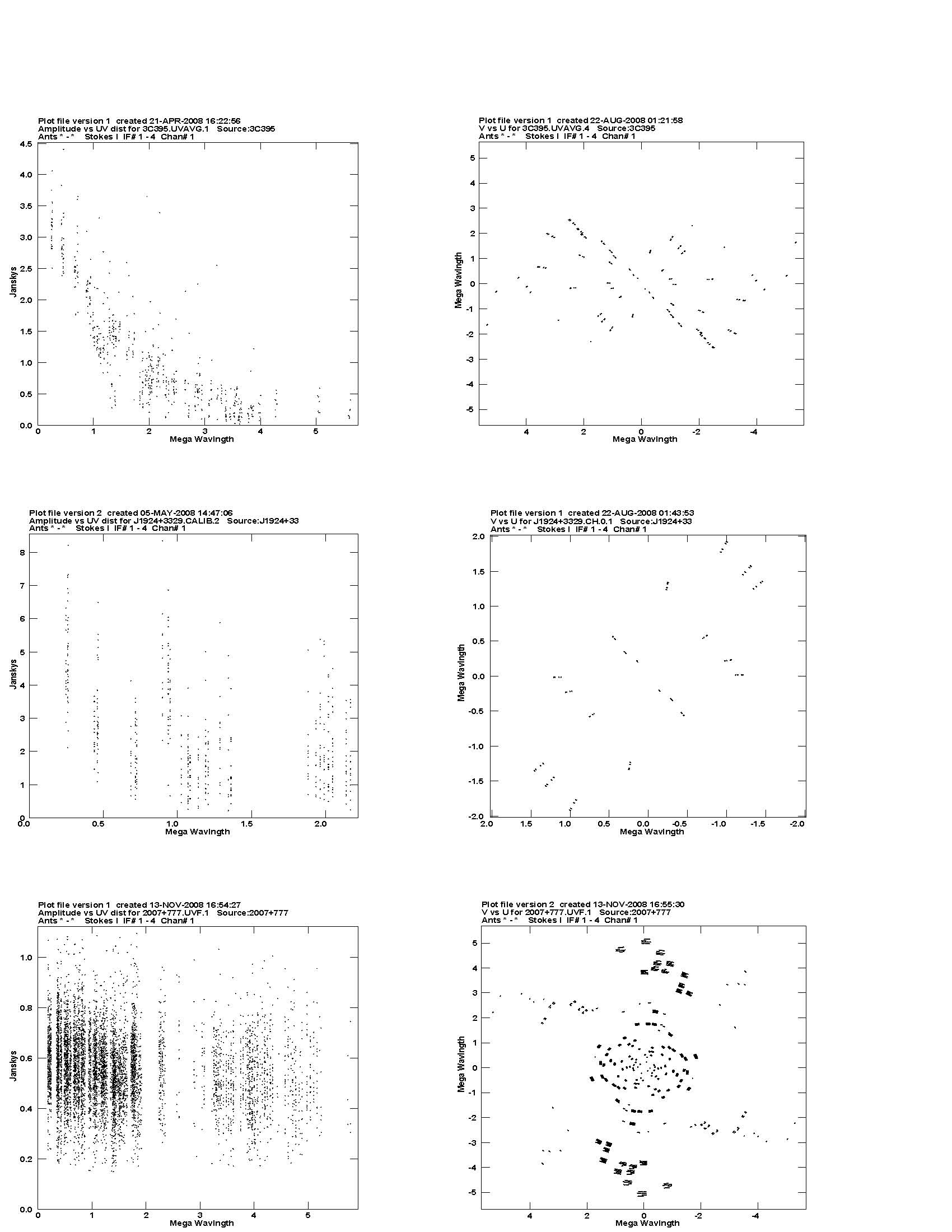} 
\end{figure}

\begin{figure}
\caption{Plots of flux density vs UV distance and UV plots for J2022+6136, J2038+5119 (3C418), J2202+4216 (BL Lac)} 
\includegraphics[scale=0.9]{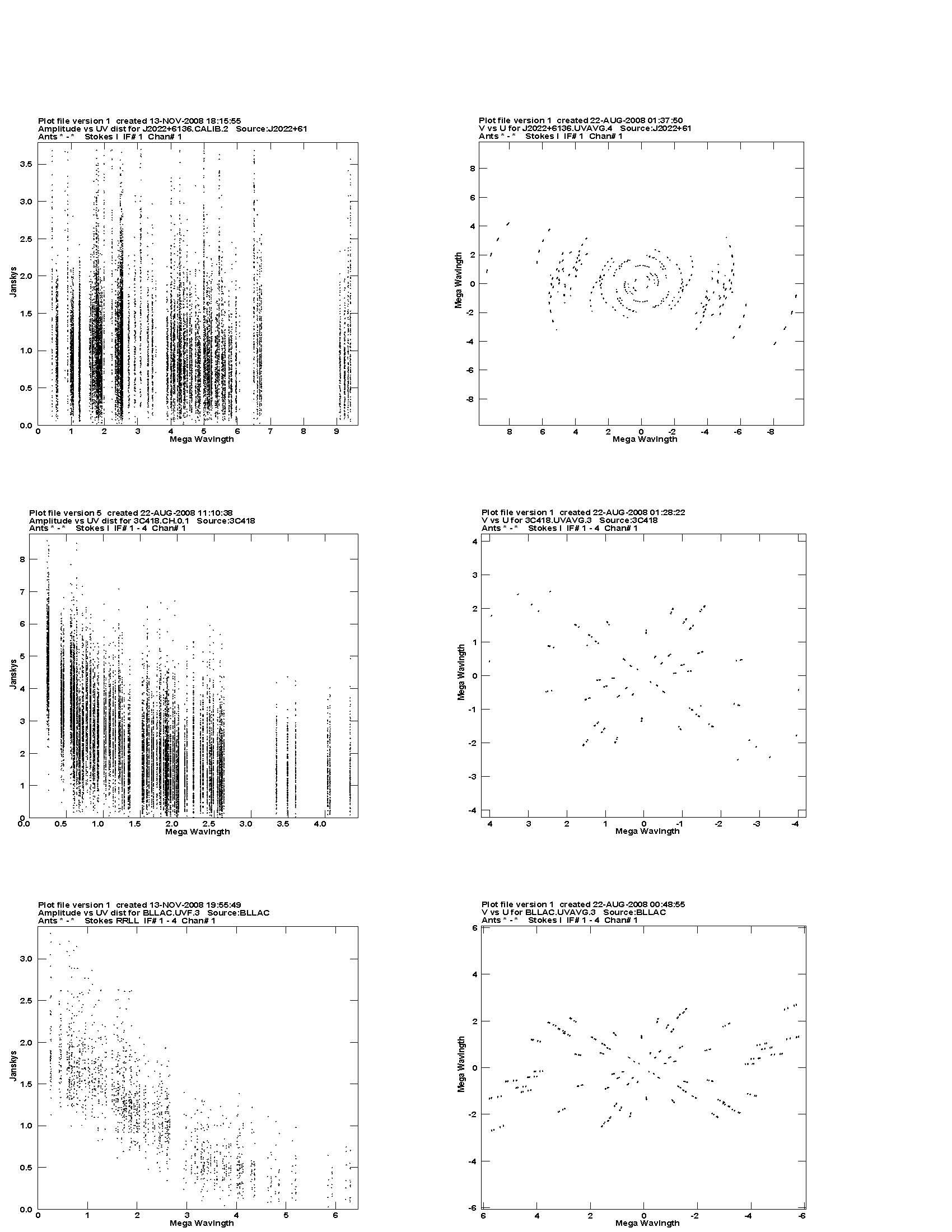} 
\end{figure}

\begin{figure}
\caption{Plots of flux density vs UV distance and UV plots for J2253+1608 (3C454.3), J2327+0940, J2330+11} 
\includegraphics[scale=0.9]{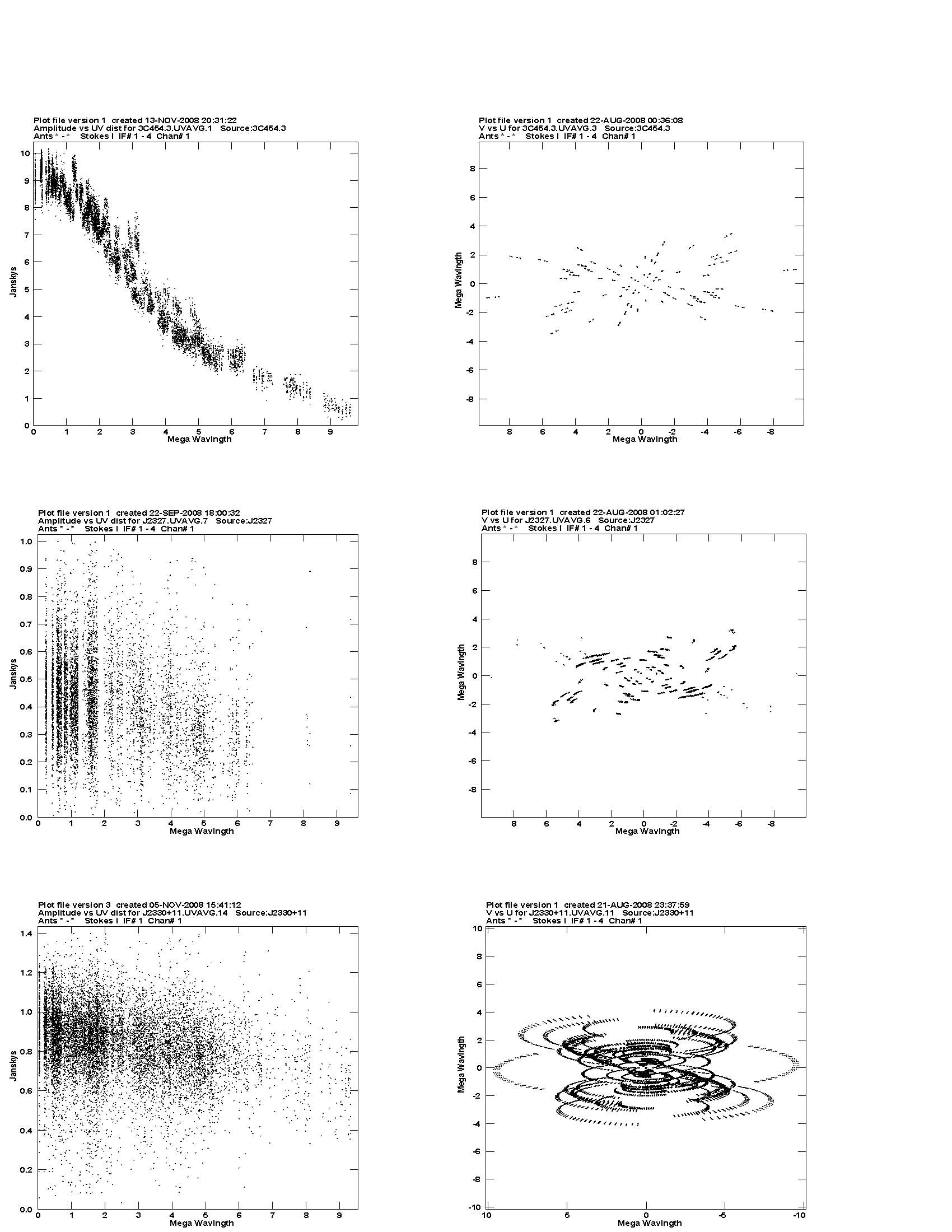} 
\end{figure}


\end{document}